2 March 2013

# Climate variability according to triple saros gravity cycles

William R. Livingston
Olympia, WA  98501, USA

**ABSTRACT:** I describe a climate model which corresponds directly to eclipse cycles. The theory is based upon a similarity between the 54-year triple saros eclipse period and the periodicity of drought. I argue that eclipse shadows are an indication of gravity cycles, and that variable lunar gravitation is the most significant aspect of the eclipse process. I reinforce the idea that lunar gravitational forcing has a profound effect on the water vapor in Earth's atmosphere, and can affect the density and location of clouds. I explore the possibility that decadal variability of ocean surface levels may be explained by triple saros gravity cycles. I point out that lunar gravitation was excluded from the most significant climate report of 2007, and that climate data contradictions have been overlooked by researchers. I focus on the value of data that has not been aggregated into global averages. I touch upon the history of global warming, and I offer predictions based upon 54-year climate periodicity. I am hopeful that readers will respond to these ideas so that the best of them can be pursued and the worst of them can be discarded. Please send correspondence to 54.year.cycle@gmail.com. This essay will be rewritten after reader responses have been received.

## 1  Introduction

It has been known for centuries that an 18-year eclipse period governs the physical relationship between the Sun, the Earth, and the Moon, and that three of these periods connected end to end synchronizes very well with the daily rotation of the Earth.  This 54-year synchronicity period is called triple saros.

I will describe a climate model that corresponds to the triple saros time period.  For this to be understood, eclipse cycles must be recognized as gravity cycles.  I propose that recognition of a 54-year gravity cycle may improve the interpretation of historical climate data, and lead the way towards improved long-range predictions of ocean surface levels, ocean temperature, atmospheric temperature, and local climate anomalies including monsoons and droughts.  I offer this climate model as an alternative to the anthropogenic global warming and climate change theories that have gained such widespread acceptance in recent years.

My search for proof that the Earth's climate is healthy led me to investigate the Moon as a possible climate catalyst, and I have noticed a similarity between the 54-year triple saros period and the periodicity of drought.  The following graphic illustrates the comparison.  While there may be a lack of consensus regarding the dates shown, these numbers are a good starting point for understanding the concept I am proposing.

Climate Variability According to Triple Saros Gravity Cycles

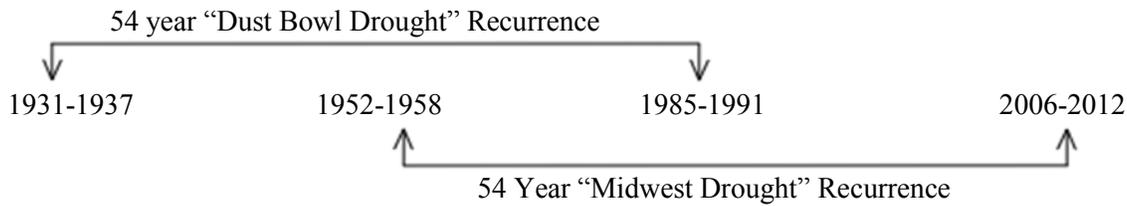

54 year "Dust Bowl Drought" Recurrence

1931-1937        1952-1958        1985-1991        2006-2012

54 Year "Midwest Drought" Recurrence

## 2   Triple saros gravity defined

The duration of a triple saros period is precisely 19,755.96 days which is the amount of time required for a lunar eclipse shadow to revisit a specific location on the surface of the Earth.  Each time the shadow touches the Earth, it is a reminder that the 54-year gravitational relationship exists.  If we can accept that gravitation is the most significant component of eclipse cycles, we can begin to focus more attention on the cyclical, physical phenomena caused by the lunar orbit.

Many researchers have focused attention on an 18.6-year periodicity of ocean tides without recognizing that the 18-year saros cycle does not synchronize with the Earth's rotation.  While the Earth's core does share an 18-year gravitational relationship with the Moon, the Earth's surface is governed by a 54-year relationship.   Therefore, any physical phenomena that might be caused by variable gravitation such as droughts, earthquakes, high ocean surface levels and cyclonic storms should be more predictable on a 54-year time scale if long-range prediction is possible.

## 3   Extrapolation of classic tidal theory

If the Moon plays a significant role in the regulation of ocean tides, it is logical that water vapor in Earth's atmosphere is also regulated by the same gravitational mechanism.  If so, the density and location of clouds must comply with the 54-year triple saros cycle.  The next logical conclusion would be that the Moon must also control heating and cooling of the Earth to some extent because clouds regulate the quantity of solar energy that hits the surface of the Earth at any given location. Clouds during winter months help to retain Earth's heat. Clouds during summer months have a cooling effect as they reflect incoming solar radiation.

Precipitation corresponds directly to cloud density. A lack of precipitation corresponds to drought.  If Earth's cloud cover is manipulated by the Moon, the Moon is partly responsible for the regulation of Earth's climate.

## 4   Declination of ocean tides

It is traditionally accepted that the Moon's orbital position affects the Earth in a generalized way, and that the Moon causes tides by dragging ocean bulges from east to west - but shouldn't we also assume that tidal bulges can be dragged from north to south as the declination of the Moon rises and falls?

Because oceans are liquid, we generally conclude that tidal bulges dissipate when the Moon is no longer overhead.  But if tidal bulges possess an elasticity or momentum that causes them to stay where the Moon places them, how might our understanding of the tides change?  It is obvious that tides rise and fall on a daily time scale, but can we definitively say that they don't also drift from region to region including north to south on a triple saros time scale of 54 years?

## 5   Climate change: the missing factor

Those who believe that humans are causing the Earth to overheat and ocean levels to rise will likely argue that all natural tidal and temperature mechanisms have been factored into the climate change calculations, but I believe that they have not. Proof of this deficiency can be found in the very document that caused worldwide fear of global warming.  The report was released by the United Nations several years ago, entitled: "Climate Change 2007: The Physical Science Basis".

The report was intended as a warning that the surface of our oceans was rising at a dangerously fast rate. But the report provides limited information regarding ways that ocean surface levels fluctuate naturally. It also lacks a description of the technique that allows scientists to differentiate between the natural fluctuations of a healthy planet and the unnatural fluctuations caused by pollution in the atmosphere.

The report never mentions the Moon as a causal factor in ocean surface variability. In fact, the Moon is mentioned only twice in the entire report, and never in relation to the tides, which is fairly remarkable considering that there may be unanimous agreement within the scientific community that the Moon plays a major role in tidal fluctuation.

# 6   Unpredictable decadal variability

The editors of the United Nations report say that certain types of climate predictions are impossible because of limited historical data. They say definitively that the surface levels of Earth's oceans are rising, while simultaneously indicating that tides fluctuate in ways that researchers do not understand. The term they use to describe the mystery is "decadal variability". The report clearly states that "decadal variability in sea level rise remains poorly understood". If this is true, why should we favor the opinion that sea levels are rising too fast?

If we believe that the Moon causes tides in our oceans, we should consider the possibility that the mysterious decadal variability in ocean surface elevation might be regulated by the triple saros gravity cycle. If this cycle also influences the density and location of clouds, monsoons might be analogous to "high tide" in Earth's atmosphere, and drought might be analogous to "low tide". If Earth's atmosphere is governed by the triple saros gravity cycle, we should be able to track droughts and predict them to some extent; and the same would be true regarding the long-term ocean surface elevation changes which are currently called "decadal variability".

# 7   Sea level contradictions

While the United Nations report of 2007 makes many generalized statements about the rise of ocean surface levels, the report also says that ocean levels are declining. As of 2007, sea levels seem to be rising on one coast of Australia while levels seem to be falling on the other coast. If the editors of the report cannot explain this conflict, why should we accept their generalized conclusion that the quantity of water in our oceans is increasing? Why should we not conclude that the water is moving from place to place?

Conflicting information also surrounds an island near Fiji called Funafuti which was made famous by a documentary film called "An Inconvenient Truth". Funafuti was also mentioned in the United Nations report of 2007 because the island had experienced some flooding. The U.N. report cites a research paper which estimates the sea level rise at Funafuti to be 2.0 millimeters per year plus or minus 1.7 millimeters per year. This means that the actual rise could be 0.3 mm or 3.7 mm. The difference between these two numbers is more than 1200%, indicating that the researchers do not know if the ocean is rising or if the island is sinking. Their uncertainty is compounded by a very short tidal record which, at the time of the U.N. report, consisted of only 14 years of high quality data. Fourteen years is only 25% of a 54-year Triple Saros period.

# 8   Climate prediction: regional vs. global

If we can learn to use the triple saros gravity cycle to track droughts as they drift across continents, and track decadal variability of the tides as they drift across oceans, we might finally be able to recognize droughts and ocean bulges as regional events rather than global events. Within this scenario we must avoid calculations that represent global averages, and we must instead focus on the raw data from each individual tide gauge, rain gauge, and temperature gauge for no less than 54 years. We should also consider that the occurrence of a drought in one place may correspond to a wetter than normal monsoon somewhere else on Earth, even in the middle of oceans where they may not affect us. If we can manage to stop worrying about ocean surface levels, and instead begin to see fluctuations and bulges as natural occurrences, we might find that increased surface levels in one part of an ocean coincides perfectly with decreased elevation somewhere else, even in the opposite hemisphere.

## 9  Global warming history

In regard to global warming, the oldest continuous temperature record is from central England. The record seems to indicate that the mean temperature of the Earth has increased by one degree in three hundred and fifty years, a third of a degree per century. But if we look at the temperature chart without comparing it to world history we won't see that the temperature record begins just after Europe had experienced the worst effects of what has come to be known as "The Little Ice Age". So back then, Europe was colder and wetter. Today, Europe seems to be warmer and dryer. The following is my triple saros version of what happened.

## 10  Theoretical speculation

The duration of a triple saros period is 54 years plus about one month. This means that a period that begins in January will end in February, 54 years later; and a period that begins in February will end in March, and so on. Therefore, the time span separating two periods that begin in January is nearly 595 years. If 595 is the number of years separating two cold periods, we might conclude that a hot period would occur at the half way point, 298 years into the cycle.

The coldest year on record in central England was 1695. If we theorize that 1695 was the coldest year for the entire Earth, warming should peak 298 years later in 1993. In close agreement with this hypothesis is a climate data set called HadSST3 from The UK's national weather service. HadSST3 indicates that the warmest year for Earth's oceans was 1998, which is within just five years of the 1993 approximation mentioned above.

## 11  Climate: fear and acceptance

Natural climate drift has been occurring forever. Without leap years and leap seconds, winter would eventually occur in July. It is the job of our time keepers to make periodic adjustments to the calendar so that we never get confused as to the time that our crops should be planted. The result of the constant adjustment of the calendar is that we have grown to expect that next year's weather should be very similar to this year's weather; and when the climate behaves badly, we worry.

Fear is a necessary thing – it is one of the instincts that keep us alive. Caring about our environment is also an important quality. There was a time not too long ago when each family in America had their own garbage dump behind the house, and our rivers near cities were polluted with sewage and trash. It is good that we are willing to make changes in order to clean up the Earth. If it is possible for us to produce less $CO_2$ pollution I think we should do so, but I am not convinced that $CO_2$ is the cause of global warming.

I am not the person who will prove that triple saros gravity is or is not the regulator of Earth's climate, but for the time being I will favor the idea that the warming of the Earth is part of a natural process. Even within the United Nations report of 2007 there is a glimmer of hope that humans are not the cause of global warming. The following quote is one example of this.

"...the present imbalance might be a rapid short term adjustment, which will diminish during coming decades".

This kind of hopeful statement is rare in the 1000 page report, but it is an indicator that editors of the report know that they could be wrong about anthropogenic global warming.

## 12  SST perspective

1944 held the record for the warmest sea surface temperatures until 1997. The warmest year for sea surface temperatures was 1998. The difference between 1944 and 1998 is 54 years.

## 13  Conclusion and predictions

Based upon triple saros gravity cycles I predict that the Earth is currently experiencing a cooling phase.

The next peak year for sea surface temperatures should be 2052. Sea surface temperatures of that year should be very similar to SSTs of 1944.

The "Dust Bowl" drought will revisit North America from 2039 to 2045. The "Midwest" drought will revisit North America from 2060 to 2066.

Climate Variability According to Triple Saros Gravity Cycles

Below is an image of worldwide temperatures for February 2013 from the UK's national weather service. Blue colors signify temperatures that are colder than average. Red colors signify temperatures that are warmer than average. Triple saros gravitation would cause the various colors to change location and intensity according to a 54-year cycle.

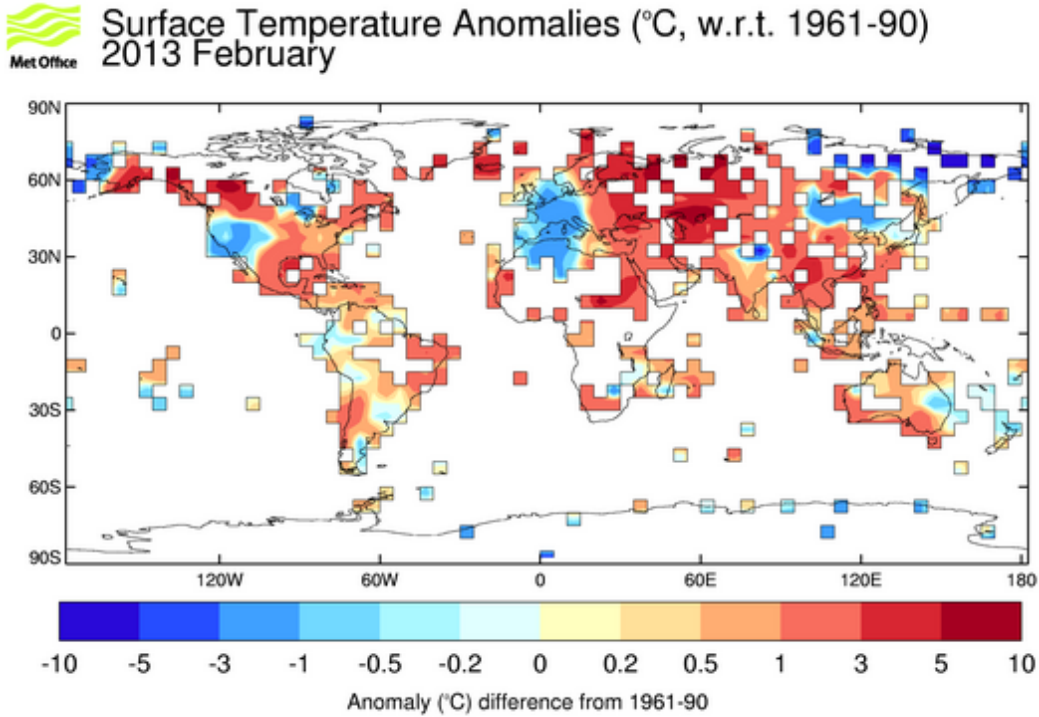

Contains public sector information licensed under the Open Government License v1.0

IPCC, 2007: Climate Change 2007: The Physical Science Basis. Contribution of Working Group I to the Fourth Assessment Report of the Intergovernmental Panel on Climate Change [Solomon, S., D. Qin, M. Manning, Z. Chen, M. Marquis, K.B. Averyt, M. Tignor and H.L. Miller (eds.)]. Cambridge University Press, Cambridge, United Kingdom and New York, NY, USA, 996 pp.

Kennedy J.J., Rayner, N.A., Smith, R.O., Saunby, M. and Parker, D.E. (2011b). Reassessing biases and other uncertainties in sea-surface temperature observations since 1850 part 1: measurement and sampling errors. J. Geophys. Res., 116, D14103, doi:10.1029/2010JD015218 (PDF 1Mb)

Kennedy J.J., Rayner, N.A., Smith, R.O., Saunby, M. and Parker, D.E. (2011c). Reassessing biases and other uncertainties in sea-surface temperature observations since 1850 part 2: biases and homogenisation. J. Geophys. Res., 116, D14104, doi:10.1029/2010JD015220 (PDF 1Mb)

Time series data file for sea surface temperatures: HadSST3_annual_globe_ts.txt